# Investigation of the Influence of a Novel Dimensionless Parameter - the Centrifugal Work Number(*CW*), on Spanwise Rotating channel Low-speed Compressible Flow

Junxin Che[1,2], Ruquan You[1,2,*], Fei Zeng[3], Haiwang Li[1,2], Wenbin Chen[3], Zhi Tao[1,2]

1 National Key Laboratory of Science and Technology on Aero Engines Aero-thermodynamics, Beihang University, Beijing 100191, China

2 Research Institute of Aero-Engine, Beihang University, Beijing 100191, China

3 Hunan Key Laboratory of Turbomachinery on Small and Medium Aero-Engine, jianguol, Zhuzhou 412002, China

**Abstract**

In the study of rotating channel flow, the key dimensionless parameters typically include the Reynolds number, rotation number, Prandtl number and buoyancy number. Our research focused on comparing the flow characteristics between the enlarged model, analyzed under the rotating similarity theory, and the original channel flow. Significantly different flow behaviors were observed between these two cases. Through theoretical derivation and dimensional analysis, we identified a new significant parameter - the centrifugal work number (*CW*). This parameter characterizes the ratio of centrifugal work to gas enthalpy in the rotating channel and plays a crucial role in measuring the compressibility of fluids within the rotating channel. Additionally, we utilized large eddy simulation(LES) to validate the impact of the centrifugal work ratio on the flow state of the rotating channel, thus enhancing the similarity theory of rotating channel compressible flow.

## 1 Introduction

In this paper, we introduce a novel dimensionless parameter, termed the centrifugal work number, derived from the rotational energy equation. This parameter quantifies the ratio of centrifugal force work to gas enthalpy in the flow within rotating channels. Numerical methods were employed to validate the influence of this parameter on rotating channel flow.

In the early 1950s, the need to cool turbine blades in aviation engines pushed for research on the flow inside rotating channel since the turbine inlet temperatures were nearing the limits of the blade materials.

The introduction of similarity theory and dimensionless parameters in the study of fluid flow aims to describe and analyze the characteristics of fluid motion more effectively. For rotating channel flow, the four key dimensionless parameters typically considered are the Reynolds number, rotation number, Prandtl number, and buoyancy number. We generally use inlet parameters as global parameters for similarity studies. When the walls of the rotating channel are adiabatic, we only need to consider two parameters: the Reynolds number and the rotation number.

* E-mail address: youruquan10353@buaa.edu.cn (R. You).

$$Re_{in} = \frac{\rho U_{in} D_h}{\mu} \tag{1.1}$$

$$Ro_{in} = \frac{\Omega D_h}{U_{in}} \tag{1.2}$$

Where $\rho$ is the fluid density, $U_{in}$ is the volume-averaged velocity at the inlet position, $D_h$ is the hydraulic diameter of the channel, $\mu$ is the dynamic viscosity of the fluid, and $\Omega$ is the rotational speed of the channel.

By ensuring consistency among these key dimensionless parameters, the influence of rotational forces on the flow within the channel can be appropriately accounted for.

Research on the effect of centrifugal force on the flow inside rotating channels is relatively scarce. In flow with constant density, centrifugal force does not have an independent effect on the flow because its effect is generally manifested as an enhanced radial pressure, which can be combined with the pressure term to represent effective pressure.

$$p_{eff} = p - \frac{1}{2}\rho\omega^2 r^2 \tag{1.3}$$

However, centrifugal force may have an impact on the rotating channel flow when density changes exist in the system(Johnston 1998)[1].

The Coriolis force plays an important role in the flow inside rotating channels, and its relative magnitude is represented by the rotational number. Barua(1954)[2] first introduced the rotational added force term in the control equation describing the flow in spanwise rotating channels, and predicted the flow pattern of Coriolis force-induced secondary flow. Moon(1964)[3] experimentally measured the velocity profiles distribution in the core region of a straight channel under rotational conditions, and observed an increase in boundary layer thickness near the leading side, and a decrease in boundary layer thickness near the trailing side. Kristoffersen(1993)[4] studied the effect of Coriolis force through direct numerical simulations, while Johnston(1972)[5] investigated the effect through experiments. They all found that the mean velocity profile was pushed toward the trailing side by the Coriolis force. They also discovered that rotation stabilized the flow near the leading side but disrupted the flow near the trailing side. With the advancement of modern fluid testing techniques, Liou(2003)[6] used LDV to obtain spectral analysis of flow in rotating channels. Sante(2010)[7] and Visscher(2011)[8] observed an increase in boundary layer thickness near the leading side and a decrease in boundary layer velocity near the trailing side under rotational conditions using PIV.

In recent years, research on turbulence in rotating channels has become increasingly deep. Regarding the impact of rotation on turbulence, Nakabayashi(2005)[9] and Brethouwer(2016)[10](2017)[11] found that even in high Reynolds number channels, the leading side begins to transition to laminar flow as the rotation number increases. However, at high Reynolds number and high rotation speeds, linearly unstable Tollmien-Schlichting-like waves can cause sustained bursting of turbulence. Grundestam(2008)[12] discovered that a balance between negative turbulence production mechanisms played a significant role in maintaining the relatively distinct interface between the two regions.

Experimental studies of the flow and heat transfer characteristics in rotating machinery internal cooling channel are limited by experimental conditions and testing techniques, requiring the scaling up of research models. Through dimensional analysis of the energy equation, we obtained another critical dimensionless parameter that determines the compressibility of the flow in the rotating channel and verified the effects of this dimensionless parameter.

## 2 Derivation of rotational dimensionless parameters

In a rotating channel, the energy equation that neglects radiation heat transfer is shown in expression 2.1.

$$\nabla \cdot \left[\left(\rho e + \rho \frac{V^2}{2} + p\right)\vec{V}\right] = \nabla(\lambda \nabla T) - \rho \vec{\Omega} \times \left(\vec{\Omega} \times \vec{r}\right) \cdot \vec{V} + \nabla \cdot (\vec{V} \cdot \tau_{ij}) \tag{2.1}$$

where $\nabla$ is the Hamiltonian operator, $\rho$ is the density, e is internal energy of the fluid, $\vec{V}$ is the velocity vector, $p$ denotes the pressure, $\lambda$ represents the thermal conductivity of the fluid, T represents the temperature, $\Omega$ represents rotation speed of the channel, and $\tau_{ij}$ denotes the tensor form of viscous force. In rotating channels, the impact of viscous forces on energy compared to positive pressure is much smaller, so the energy equation for steady flow neglecting the work done by viscous forces can be expressed as expression 2.2:

$$\nabla \cdot \left[\left(\rho e + \rho \frac{V^2}{2} + p\right)\vec{V}\right] = \nabla(\lambda \nabla T) - \rho \vec{\Omega} \times \left(\vec{\Omega} \times \vec{r}\right) \cdot \vec{V} \tag{2.2}$$

Simplifying and expressing the relative magnitude of total energy in terms of enthalpy and kinetic energy, there are only two ways to change the total energy of fluids in a rotating channel: heat transfer and work done by centrifugal force.

$$\nabla \cdot \left[\rho c_p T \vec{V} + \rho \frac{V^2}{2} \vec{V}\right] = \nabla(\lambda \nabla T) - \rho \vec{\Omega} \times \left(\vec{\Omega} \times \vec{r}\right) \cdot \vec{V} \tag{2.3}$$

The change in energy in a rotating channel is mainly achieved through heat transfer and work done by centrifugal force. To examine the similarity of flow, different flow fields need to be compared, and the various terms in the energy equation need to be made dimensionless. Let $c_{p0}$, $U_0$, $\rho_0$, $\theta_0$, $D_0$ and $r_0$ represent the characteristic physical quantities of specific heat at constant pressure, velocity, density, temperature, length, and rotation radius, respectively. The dimensionless quantities are represented by *.

$$c_p^* = \frac{c_p}{c_{p0}}, \quad \vec{V}^* = \frac{\vec{V}}{U_0}, \quad \rho^* = \frac{\rho}{\rho_0}, \quad T^* = \frac{T}{\theta_0}, \quad \vec{r}^* = \frac{\vec{r}}{r_0} \tag{2.4}$$

By substituting dimensionless quantities into the energy equation, expression 2.5 is obtained.

$$\left(\frac{\rho_0 c_{p0} T_0 U_0}{D_0}\right) \nabla^* \cdot \left(\rho^* c_p^* \theta^* \vec{V}^*\right) + \left(\frac{\rho_0 U_0^3}{D_0}\right) \nabla^* \cdot \left(\rho^* \frac{V^{*2}}{2} \vec{V}^*\right) = \left(\frac{\lambda T_0}{D_0^2}\right) \nabla^* (\nabla^* \theta^*) - \left(\rho_0 \Omega^2 r U_0\right) \rho^* \left[\vec{i} \times \left(\vec{i} \times \vec{r}^*\right)\right] \cdot \vec{V}^* \tag{2.5}$$

The dimensionless energy equation can be simplified and expressed as expression 2.6 after rearrangement.

$$\nabla^* \cdot \left( \rho^* c_p^* \theta^* \vec{V}^* \right) + \left( \frac{U_0^2}{c_{p0} T_0} \right) \nabla^* \cdot \left( \rho^* \frac{V^{*2}}{2} \vec{V}^* \right) =$$
$$\left( \frac{\lambda}{\rho_0 U_0 D_0 c_{p0}} \right) \nabla^* (\nabla^* \theta^*) - \left( \frac{\Omega^2 r_0 D_0}{c_{p0} T_0} \right) \rho^* \left[ \vec{i} \times \left( \vec{i} \times \vec{r}^* \right) \right] \cdot \vec{V}^* \tag{2.6}$$

The boundary conditions for the flow also need to be made dimensionless. The flow boundary conditions in a rotating channel are the same as those for stationary channels, so they will not be derived here. From the above dimensionless energy equation, it is clear that as long as the corresponding dimensionless combinations of parameters for the two flow phenomena are the same, then they are similar. These dimensionless combination coefficients are the control criteria for flow similarity.

$$\frac{U_0^2}{c_{p0} T_0} = (k-1) \frac{U_0^2}{k R_g T_0} = (k-1) Ma^2 \tag{2.7}$$

$$\frac{\lambda}{\rho_0 U_0 D_0 c_{p0}} = \frac{\mu_0}{\rho_0 U_0 D_0} \cdot \frac{\lambda}{\mu_0 c_{p0}} = \frac{1}{Re \cdot Pr} \tag{2.8}$$

The dimensionless combination parameters before the kinetic energy term and the heat transfer term are the same as those in stationary flow. This can be simplified into three key parameters: *Ma*, *Re*, and *Pr*. *Ma* represents the impact of compressibility caused by fluid velocity on the flow. Generally, when Ma is less than 0.3, the impact on the flow can be ignored.

The final term, $\frac{\Omega^2 r_0 D_0}{c_{p0} T_0}$, represents the ratio of the work done by centrifugal force to the enthalpy of the fluid, which characterizes the relative magnitude of work done by centrifugal force. The dimensionless parameters can be expressed in terms of commonly used parameters, which can be simplified as follows:

$$\frac{\Omega^2 r_0 D_0}{c_{p0} T_0} = \frac{\frac{\Omega^2 D_0^2}{U_0^2} \frac{r_0}{D_0}}{\frac{1}{(k-1) Ma^2}} = (k-1) Ro^2 \frac{r_0}{D_0} Ma^2 \tag{2.9}$$

The dimensionless parameter can be expressed as a combination of Rotation number *Ro*, ratio of rotation radius to hydraulic diameter $r/D_h$, and Mach number *Ma*. Although a new dimensionless parameter has been obtained in the energy equation, further verification is still needed to determine the magnitude of its impact on the flow.

## 3 Effect of *CW* on flow compressibility

From the perspective of the effects, the impact of the work done by centrifugal force on the flow is mainly achieved by compressing the fluid through the work done along the path. Using the derived results, we can further compare the relative sizes of the work done by centrifugal force and the kinetic energy term within the internal cooling channel. The relative size of the gas kinetic energy term is represented as $(k-1)Ma^2$, while the work done by centrifugal force term is represented as $(k-1)Ro^2 \frac{r_0}{D_h} Ma^2$. Taking a typical operating condition of the turbine internal cooling channel in aero-engines as a reference, the range of $Ro$ is usually between 0.3 and 0.4, while the range of $r/D_h$ is between 70 and 100. Therefore, the range of $Ro^2 \frac{r_0}{D_h}$ is between 6.3 and 16. In other words, under the same $Ma$, the impact of the work done by centrifugal force on gas energy is much stronger than the impact of kinetic energy. In order to simplify the expression, we will refer to the parameter that characterizes the impact of the work done by centrifugal force on the flow as the Centrifugal Work number, denoted as $CW$:

$$CW = Ro^2 \frac{r}{D_h} Ma^2 \tag{3.1}$$

Here, we simplify the flow in the rotating channel into a one-dimensional radial flow, and then derive the effect of centrifugal work on fluid compressibility.

The one-dimensional momentum equation in the $z$-direction of the channel, neglecting frictional forces, can be expressed as:

$$w \frac{dw}{dz} = \Omega^2 r - \frac{1}{\rho} \frac{dp}{dz} \tag{3.2}$$

The speed of sound of the idea gas can be expressed as:

$$c^2 = \frac{dp}{d\rho} \tag{3.3}$$

By substituting the momentum equation 3.4, we obtain:

$$w^2 \frac{dw}{w} = \Omega^2 r dz - \frac{d\rho}{\rho} c^2$$

$$Ma^2 \frac{dw}{w} = \frac{\Omega^2 r dz}{c^2} - \frac{d\rho}{\rho} \tag{3.4}$$

$$\frac{d\rho}{\rho} = CW \frac{dz}{D_h} - Ma^2 \frac{dw}{w}$$

It can be seen that the change in fluid density is mainly affected by the $CW$ and $Ma$. In order to further explore the effect of the $CW$ on the compressibility of the fluid, let's simplify the equation further:

$$\frac{d\rho}{\rho} = CW \frac{dz}{D_h} - Ma^2 \frac{d\left(\frac{\dot{m}}{\rho A}\right)}{\frac{\dot{m}}{\rho A}} \tag{3.5}$$

Assuming there is no flow out and the cross-sectional area remains constant along the duct, we can further simplify the equation:

$$\frac{d\rho}{\rho} = CW \frac{dz}{D_h} + Ma^2 \frac{d\rho}{\rho} \tag{3.6}$$

$$\frac{d\rho/\rho}{dz/D_h} = \frac{CW}{1 - Ma^2} \tag{3.7}$$

The rate of change of density per unit length is influenced by both Ma and the centrifugal work number (*CW*).

When the Mach number of the flow inside the rotating channel is less than 0.3, the effect of fluid density variation rate is about 0.1 times that of the *CW*. We define the compressible flow inside the rotating channel as a low-speed compressible flow. The equation can be simplified as follows:

$$\frac{d\rho/\rho}{dz/D_h} = CW \tag{3.8}$$

Therefore, in the low Mach number flow inside the rotating channel, the key factor determining the compressibility of the fluid is the centrifugal work. When applied to practical engineering projects, most of the flow in the rotating machinery's cooling channels is in a state of low-speed compressible flow.

The key dimensionless parameter, centrifugal work (*CW*), plays an important role in studying the flow in rotating channels. In engineering applications research, we need to pay special attention to the impact of compressibility effects generated by centrifugal force on the flow in rotating channels. Conclusions that do not consider compressibility effects and simply model the effects of Coriolis force, buoyancy force, and other effects may not reflect the real flow state.

When modeling compressible flow in rotating channels in the laboratory, we can no longer freely choose parameters such as inlet pressure and inlet temperature of the channel. Instead, we must adjust the inlet temperature and pressure to match the inlet *CW* with the actual working model of the cooling channel inside the rotating machinery. This allows us to simulate the flow inside the real rotating channel under ambient conditions in the laboratory.

# 4 Verification of the influence of *CW* on rotating channel flow
## 4.1 Test section and numerical methods

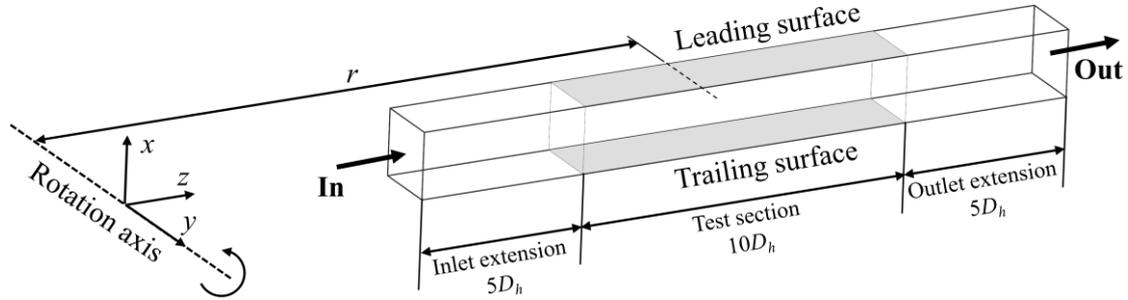

Figure 1: Sketch of physical configuration and computational model

The test section is shown in figure 1. The section of the model is a rectangular shape with an aspect ratio of 1. The length of the test section is $10D_h$, and $5D_h$ of extended sections are set at the inlet and outlet to eliminate the influence of the inlet and outlet boundary on the flow in the test section.

The boundary conditions for numerical research of the channel are set as follows:
- Inlet: Velocity; the inlet temperature is 773K or 300K;
- Wall: Adiabatic condition;

In this paper, a typical working condition of the internal cooling channel of the aero-engine turbine was selected for the research. The hydraulic diameter of the original model was *5mm*, the inlet temperature of the channel was 773K, the inlet pressure was 2MPa, the rotation speed was 16000rpm, and the inlet Reynolds number was 50000. An ideal gas was used for the research.

In order to accurately predict the flow inside the rotating channel, Large Eddy Simulation (LES) is adopted. Non-uniform grids are used to partition the cross-sections perpendicular to the flow direction. Grid refinement is applied near the walls, with a grid spacing of approximately $y^+ \approx 0.5$ for the first layer near the wall. Additionally, the maximum grid spacing in the mainstream region is approximately $\Delta y^+ \approx 20$ and $\Delta x^+ \approx 20$. The z-direction of the model is divided using a uniform grid with $\Delta z^+ \approx 40$. The computational domain is divided into 200×200×1500 elements. The computational approach for Large Eddy Simulation is presented in Table 1.

Table 1: Large eddy simulation numerical method

| Numerical Methods | | Large Eddy Simulation |
|---|---|---|
| Sub-grid Scale Model | | WALE |
| Pressure-Velocity Coupling | | Coupled |
| Spatial Discretization | Pressure | Second Order |
| | Density | Second Order Upwind |
| | Momentum | Bounded Central Differencing |
| | Energy | Bounded Central Differencing |
| Transient Formulation | | Bounded Second Order Implicit |

As shown in Figure 2, this study compares the mainstream velocity profiles obtained using Large Eddy Simulation (LES) with the ones computed by Kristoffersen (1993) [4] using Direct Numerical

Simulation (DNS) for a rotating channel. The maximum error in the mainstream velocity profiles obtained through LES is less than 2%. Therefore, it can be concluded that the numerical methods employed in our study can effectively predict the mainstream velocity profiles inside the rotating channel.

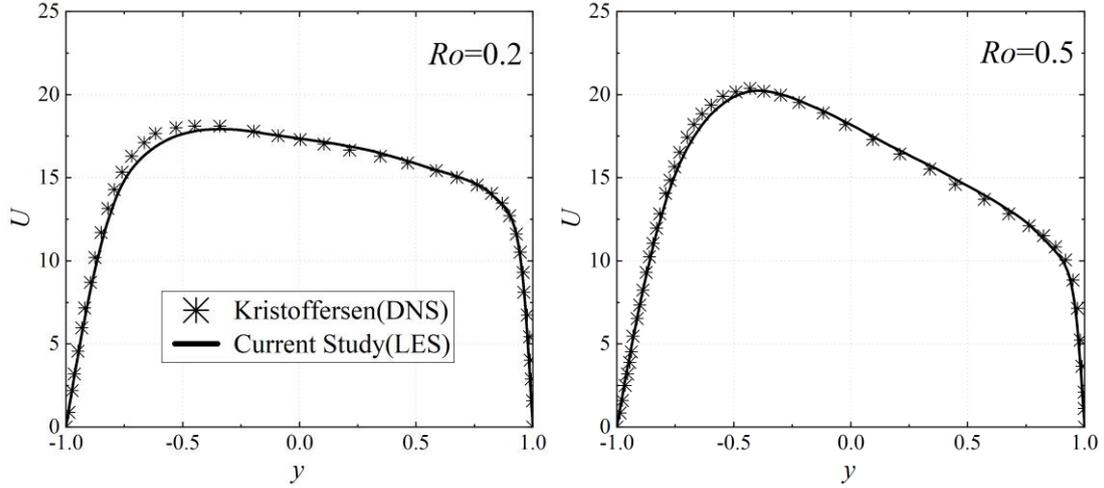

Figure 2: Comparison of mainstream velocity profiles between Large Eddy Simulation (LES) and DNS

## 4.2 Verification of dimensionless parameters for similarity theory of rotating channel incompressible flow

In a rotating adiabatic channel, modeling the flow only requires maintaining geometric and aerodynamic similarity, without considering the influence of centrifugal force. To achieve this, a consistent aspect ratio ($r/D_h = 70$), and the same aerodynamic parameters, such as Reynolds number ($Re = 50000$) and rotation number ($Ro = 0.412$), can be used. Based on this, the original model was scaled up. The parameter values for various models are presented in Table 2. During the process of model scaling, $Re$, $Ro$, and $r/D_h$ remained constant. However, during the process of model enlargement (as shown in Table 2), the constant inlet density resulted in a decrease in the inlet Mach number. This caused a gradual decrease in the work done by centrifugal force on the fluid with model enlargement, which in turn weakened the compression effect of the fluid.

Table 2: Model parameters validating the influence of inlet $CW$

| $D_h$/mm | 5 | 6 | 10 | 20 |
|---|---|---|---|---|
| $T_{in}$/K | 773 | | | |
| $p_{in}$/Pa | $2\times 10^6$ | | | |
| $n$/rpm | 16000 | 11111 | 4000 | 1000 |
| $Ma$ | $3.65\times10^{-2}$ | $2.97\times10^{-2}$ | $1.78\times10^{-2}$ | $8.91\times10^{-2}$ |
| $CW$ | $1.58\times10^{-2}$ | $1.10\times10^{-2}$ | $3.95\times10^{-2}$ | $9.88\times10^{-2}$ |

The velocity distribution of the cross-section along the channel of each hydraulic diameter is shown in figure 3, under the condition of the same inlet temperature ($T_{in}$) and inlet pressure ($P_{in}$). The velocity distribution of the cross-section along the channel of different hydraulic diameter models is shown in figure 3.

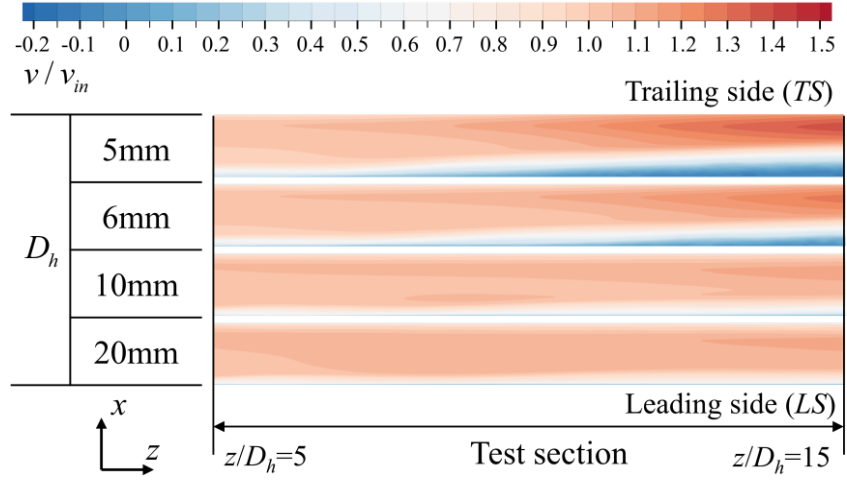

Figure 3: Comparison of cross-sectional dimensionless velocity distribution in similarity models with different hydraulic diameters.

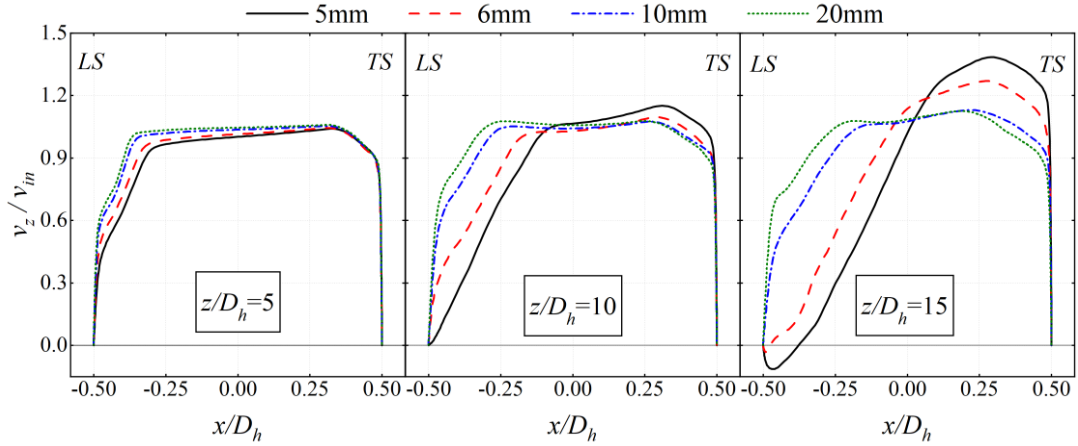

Figure 4: Streamwise velocity profiles along channels of different scaling models.

Although the same dimensionless parameters were maintained during the model scaling process, there were significant differences in the flow fields of the channel cross-sections. In an adiabatic rotating channel, due to the effect of the Coriolis force, the boundary layer near the leading side develops significantly faster than that near the trailing side, and the low-speed region near the leading side continuously expands towards the mainstream (figure 4). In the small-scale model, the boundary layer near the leading side develops rapidly while the velocity significantly decreases. As the fluid near the leading side decelerates, flow separation even occurs near the leading in small-sized channels ($D_h$ = 5$mm$ and 6$mm$). This phenomenon has not been observed in previous studies of incompressible adiabatic wall rotating channel flows.

We know that the choice of outlet boundary conditions is crucial in LES as it can have a significant impact on flow separation near the leading side. To validate if the recirculation region is influenced by the outlet boundaries, we extended the outlet by an additional 10$D_h$ while keeping the inflow parameters identical to the original model. The figure 5 and 6 below compare the velocity profiles at three locations ($z/D_h$) in the channels with $D_h$ = 5$mm$, along with the dimensionless velocity contour plot at the $x$ = 0 section. Extending the channel did not have any influence on the flow at the investigated positions, and the location of the separation region remained almost unchanged. This indicates that the observed flow separation near the leading side is physically real and is not affected by the outlet conditions.

The flow separation near the leading implies that the effect of centrifugal work on the fluid is not simply a result of compressing the fluid and reducing its velocity. A decrease in fluid velocity does cause an increase in the relative strength of the Coriolis force, resulting in a decrease and increase in flow velocity near the leading and trailing sides, respectively. However, under adiabatic wall conditions, flow separation near the leading is not observed. Indeed, during the compression process, there exists a certain effect that causes a rapid deceleration of the fluid near the leading side and can even lead to flow separation. This effect even has a greater impact on the fluid than the mere compression effect itself.

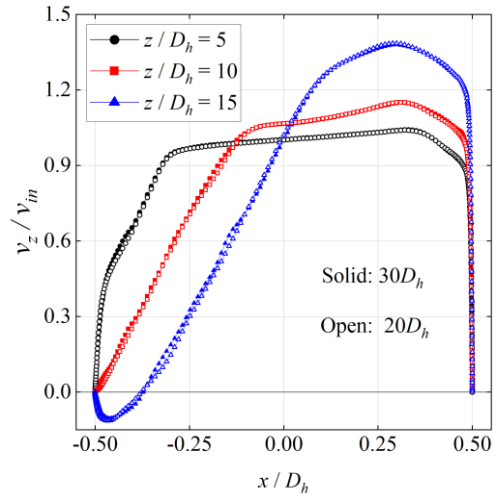

Figure 5: Velocity profiles comparison of three flow sections in models with channel lengths of $20D_h$ and $30D_h$.

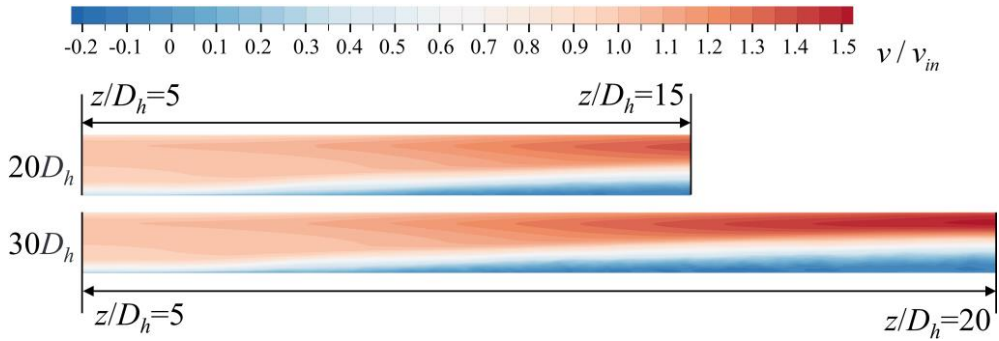

Figure 6: Velocity profiles of the midsection (x = 0) in models with channel lengths of $20D_h$ and $30D_h$.

However, in the large-scale similar model, even though the channel had the same inlet $Re$ and inlet $Ro$, the enlarged model was unable to model the state of flow in rotating channel. As the flow continuously developed, the differences in flow between different channels became more significant.

This also implies that the centrifugal work number is a critical parameter to consider in the process of similarity analysis. In the model of the internal cooling channel of the rotating machinery studied, although the inside the channel is much less than 0.3, the compressibility effect of the fluid has been quite significant and has become an important factor affecting the flow state of the fluid inside the channel.

With the enlargement of the channel model, the velocity variation gradually decreased, approaching that of incompressible flow. It can be observed that when comparing the flow inside channels with hydraulic diameters of 10*mm* and 20*mm*, significant differences are also observed at $z/D_h$=15. In the channel with a hydraulic diameter of 20mm, the velocity profiles of the fluid remain almost unchanged at the three radial positions studied. Therefore, when the centrifugal work number is less than $10^{-3}$, it can be concluded that the compressive effect of centrifugal force within the channel can be essentially negligible. In other words, when the unit length density variation rate is less than 0.1%, the flow inside the rotating channel can be approximated as incompressible flow. The numerical results show that this approximation is valid for rotating channels with hydraulic diameter lengths up to $20D_h$.

**4.3 Verification of the effect of *CW* on rotating channel compressible flow**

To verify the influence of *CW* on the similarity process of rotating channels, we conducted additional similarity studies on channels while keeping the *CW* constant (Table 3). To achieve the same *CW* at the inlet during the scaling process, the inlet pressure must be adjusted to ensure that the inlet Mach number remains consistent with the original model. This can be derived from the Reynolds number similarity and the gas state equation:

$$p_1 = \frac{p_0 \mu_{T_1}}{n \mu_{T_0}} \sqrt{\frac{T_1}{T_0}} \tag{4.1}$$

In this formula, $n$ represents the scaling ratio of the model during the modeling process. Based on this, we adjusted the rotating speed to match the rotation number Ro. As shown in Table 2, when transforming from model 0 to model 1 and model 2, the hydraulic diameter was enlarged from 5*mm* to 20*mm*, and the corresponding inlet pressure was reduced to achieve the matching of the inlet Mach number, thus obtaining a series of similarity results where the inlet Re, Ro, *CW*, and r/$D_h$ all remained consistent. Model 3 and 4 considered the use of ambient temperature gas in laboratory experiments, and the inlet temperature was changed to 300K while still keeping the above key parameters consistent.

Table 3: Model parameters for different hydraulic diameter models with the same *CW*

| model | 0 | 1 | 2 | 3 | 4 |
|---|---|---|---|---|---|
| $D_h$/mm | 5 | 10 | 20 | 10 | 20 |
| $T_{in}$/K | 773 | 773 | 773 | 300 | 300 |
| $p_{in}$/Pa | $2 \times 10^6$ | $1 \times 10^6$ | $5 \times 10^5$ | $6.23 \times 10^5$ | $3.15 \times 10^5$ |
| $n$/rpm | 16000 | 8000 | 4000 | 4983 | 2492 |
| Ma | \multicolumn{5}{c}{$3.65 \times 10^{-2}$} |
| CW | \multicolumn{5}{c}{$1.58 \times 10^{-2}$} |

When keeping the inlet Mach number the same, the nondimensional velocity field in the middle cross-section of the channel is shown in figure 7. After introducing new nondimensional parameters, the development status of the flow along the channel is almost identical to that of the original model. The velocity profile along the spanwise cross-section remains similar to that of the original model after amplification (Figure 8). In other words, after introducing the nondimensional parameter of the centrifugal work term, that is, keeping the inlet *CW* consistent, the flow characteristics of the channel are consistent with those of the similar model. Therefore, it can be verified that the key nondimensional parameter *CW* derived from the energy equation is the key nondimensional parameter that needs to be consistent in the study of rotational similarity.

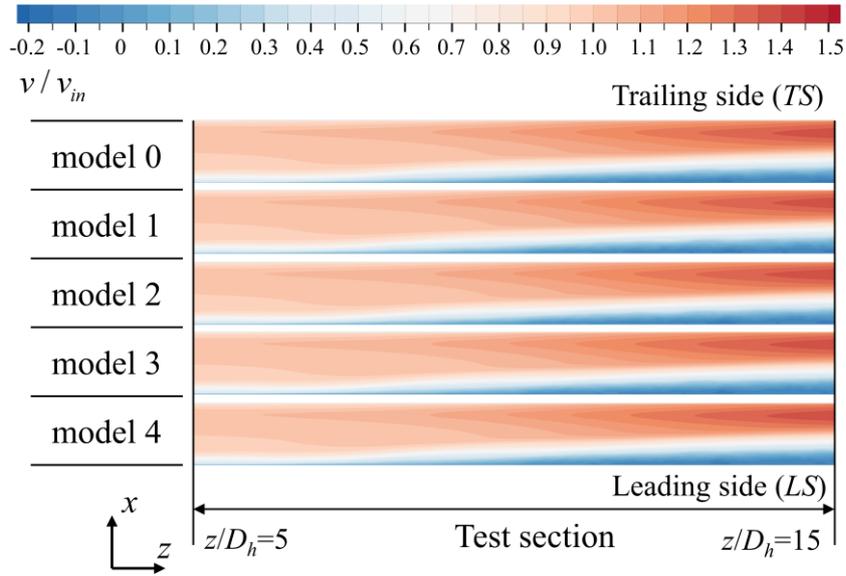

Figure 7: Distribution of dimensionless velocity field in cross-sections of similarity models with consistent inlet *CW*.

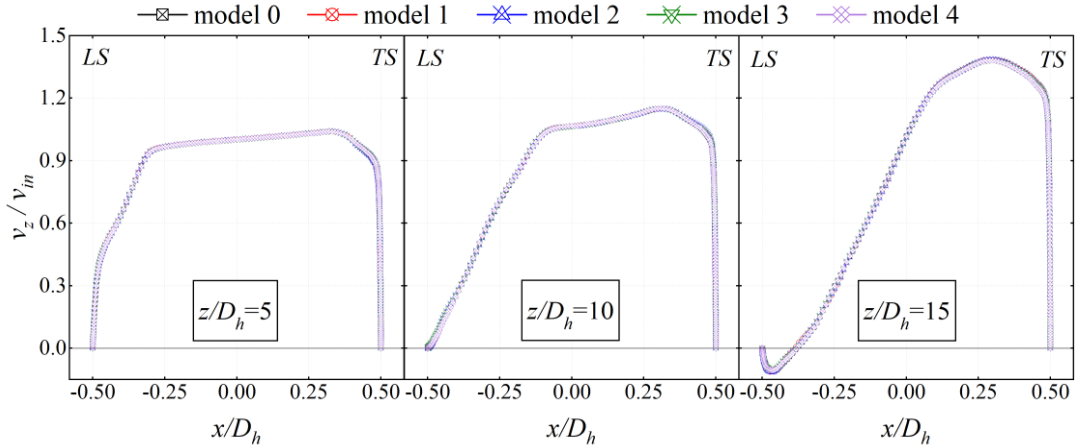

Figure 8: Streamwise velocity profiles along channels of different scaling models with consistent inlet CW.

# 5 Conclusion

This paper employs dimensional analysis and theoretical derivation to introduce a significant dimensionless parameter, known as the centrifugal work number (*CW*).

$$CW = Ro^2 \frac{r}{D_h} Ma^2 \qquad (5.1)$$

The *CW* represents the ratio of centrifugal work to gas enthalpy and plays a crucial role in determining the compressibility of flow within rotating channels. In typical internal cooling channels of rotating machinery, the compression effect of the flow cannot be disregarded due to the substantial value of the centrifugal work number. Consequently, the similarity rules derived from the incompressible Navier-Stokes equations (maintaining geometric similarity, as well as identical values of *Re*, *Ro*, $r/D_h$, and *Buo*) fail to adequately describe the flow state within these channels. For compressible flow in rotating channels, matching the centrifugal work number becomes an essential requirement.

Through theoretical derivation, we establish the relationship between flow compressibility and the centrifugal work number in a low-speed channel.

$$\frac{d\rho/\rho}{dz/D_h} = CW \tag{5.2}$$

By analyzing numerical results, we determine that when the unit length density variation rate is less than 0.1% (i.e., $CW < 0.001$), the fluid inside the channel can be approximately treated as incompressible, simplifying the flow problem.

In the study of turbulent flow using LES, flow separation near the leading side can introduce additional complexities and challenges to the study of rotating channel flows. It highlights the importance of considering factors such as $CW$ and flow conditions when analyzing and predicting flow behavior. This new observation opens up opportunities for further investigation and understanding of the underlying physics involved in these flows.

In summary, this research provides valuable insights into the understanding of centrifugal forces in fluid dynamics and offers practical guidance for the design and optimization of flow systems within rotating channels.

## Declaration of Interests

The authors report no conflict of interest.

## Acknowledgments

This work was supported by National Science Fund for Distinguished Young Scholars [No. 52225602];

This work was supported by Beijing Municipal Natural Science Foundation [No. 3222034];

This work was sponsored by Beijing Nova Program.